\documentclass[journal=jctcce,manuscript=article]{achemso}

\usepackage[version=4]{mhchem} 
\usepackage{amsfonts}
\usepackage{amssymb,amsmath}
\usepackage[linesnumbered,boxed]{algorithm2e}
\usepackage{enumerate}
\usepackage{bbm}
\usepackage{color}
\usepackage{braket}
\usepackage{subfigure}
\usepackage{subcaption}
\usepackage{graphics}

\setkeys{acs}{maxauthors = 0} 

\newcommand{\rGPU}{\mathrm{GPU}}

\author{Chunyang Xiang}
\affiliation{State Key Lab of Processors, Institute of Computing Technology, Chinese Academy of Sciences}
\alsoaffiliation{University of Chinese Academy of Sciences, Beijing, China}
\author{Weile Jia}
\affiliation {State Key Lab of Processors, Institute of Computing Technology, Chinese Academy of Sciences}
\author{Wei-Hai Fang}
\author{Zhendong Li}\email{zhendongli@bnu.edu.cn}
\affiliation{Key Laboratory of Theoretical and Computational Photochemistry, Ministry of Education, College of Chemistry, Beijing Normal University, Beijing 100875, China }

\title[\texttt{achemso} demonstration]
{A distributed multi-GPU \emph{ab initio} density matrix renormalization group algorithm
with applications to the P-cluster of nitrogenase}

\begin{document}

\begin{tocentry}

\includegraphics{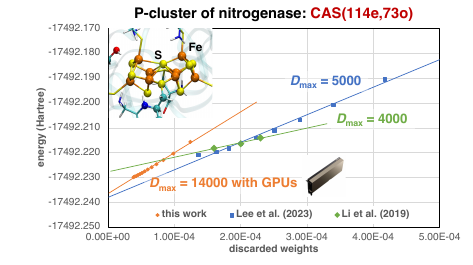}




\end{tocentry}

\begin{abstract}
The presence of many degenerate $d/f$ orbitals makes polynuclear transition metal 
compounds such as iron-sulfur clusters in nitrogenase challenging for state-of-the-art 
quantum chemistry methods. To address this challenge, we present the first 
distributed multi-GPU (Graphics Processing Unit) 
\emph{ab initio} density matrix renormalization (DMRG) algorithm,
suitable for modern high-performance computing (HPC) infrastructures.
The central idea is to parallelize the most computationally intensive part - the multiplication of $O(K^2)$ operators with a trial wavefunction, where $K$ is the number of spatial orbitals, 
by combining operator parallelism for distributing the workload 
with a batched algorithm for performing contractions on GPU.
With this new implementation, we are able to reach an unprecedentedly
large bond dimension $D=14000$ on 48 GPUs (NVIDIA A100 80 GB SXM)
for an active space model (114 electrons in 73 active orbitals) of the P-cluster, which is nearly three times larger than the bond dimensions reported in previous DMRG calculations for the same system using only CPUs. 
\end{abstract}

\section{Introduction}\label{introduction}
The density matrix renormalization group (DMRG) algorithm\cite{white1992density,white1993density} 
is a powerful numerical tool initially invented for computational study of strongly correlated
one-dimensional systems. Its adaptation to quantum chemistry often referred as 
\emph{ab initio} DMRG\cite{white1999ab,daul2000full,mitrushenkov2001quantum,chan2002highly,chan2003exact,
legeza2003controlling,legeza2003optimizing,
chan2004algorithm,mitrushenkov2003quantum,chan2004state,chan2005density,
moritz2006construction,hachmann2006multireference,marti2008density,ghosh2008orbital,chan2008density,
zgid2008obtaining,luo2010optimizing,marti2011new,kurashige2011second,
sharma2012spin,chan2012low,wouters2012longitudinal,mizukami2012more,
kurashige2013entangled,sharma2014low,wouters2014density,wouters2014chemps2,
fertitta2014investigation,knecht2014communication,szalay2015tensor,
yanai2015density,olivares2015ab,baiardi2020density,cheng2022post,ma2022density}. 
has made the study of transition metal complexes with several metal centers become possible,
where the traditional full configuration interaction (FCI)
is usually limited to one or two transition metal centers. 
Typical examples include the applications to 
the oxygen-evolving complex\cite{kurashige2013entangled} \ce{[Mn4CaO5]},
the iron-sulfur clusters with \ce{[Fe2S2]} and \ce{[Fe4S4]} cores\cite{sharma2014low},
the P-cluster \ce{[Fe8S7]} and FeMoco \ce{[Fe7MoS9C]} of nitrogenase
with eight transition metal centers\cite{li2019electronic,li2019electronic2,lee2023evaluating}.
It should be emphasized that the entanglement structure between orbitals
in these molecules is more complex than that in one-dimensional systems.
As a result, while for one-dimensional systems the ground state can be well captured 
by a finite bond dimension $D$ in the underlying variational wavefunction, 
independent of the system size\cite{eisert2010colloquium}, 
the variational energy as a function of the bond dimension $D$ converges
much more slowly for these complicated systems. This means that
the bond dimension required to reach certain accuracy (e.g. 1 milli-Hartree per metal)
for these systems needs to
increase as the system size increases.
In view of the computational scaling of \emph{ab initio} DMRG, which is $O(K^3D^3+K^4D^2)$
with $K$ being the number of spatial orbitals, this presents a
huge computational challenge, which limits the typical value of $D$
to several thousands.

{
In view of such challenges, the major goal of the present work is to develop a new parallelization algorithm for \emph{ab initio} DMRG, 
which can take advantage of modern heterogeneous high-performance
computing (HPC) infrastructures. To date, Graphics Processing Unit (GPU) 
has become the most common accelerating device in 
HPC systems, for example, seven of the top ten supercomputers on the top 500 list~\cite{top500} are equipped with GPUs. The many-core architecture of GPU is specially designed for compute-bound and memory-bound tasks such as matrix-matrix multiplication (GEMM) and fast Fourier transformation (FFT), respectively. As a result, GPU has been widely adopted in computational chemistry and physics packages, e.g., TeraChem~\cite{ufimtsev2008graphical,seritan2021terachem}, GAMESS\cite{zahariev2023general}, FermiONs++\cite{kussmann2017hybrid}, VASP~\cite{hacene2012accelerating,hutchinson2012vasp}, PWmat~\cite{JIA2013,JIA2013102}, Quantum Espresso~\cite{RomeroJoshua2018}, BerkeleyGW~\cite{gb2020_berkeleygw}, etc.
In addition, recent developments of Tensor Cores~\cite{tensorCore} 
have significantly improved the computational capability of GPUs,
in particular, GEMM in double precision can be further accelerated
on NVIDIA A100 or H100 GPUs.}

Effectively utilizing the power of modern heterogeneous HPC infrastructures
is nontrivial for \emph{ab initio} DMRG, whose parallelization is less
straightforward due to the complexity of the algorithm itself and
the data structure involved. Different parallelization 
strategies need to be combined to achieve good performance.
The first parallelization strategy developed by Chan\cite{chan2004algorithm},
which will be referred as operator parallelism, is based on 
a partition of normal/complementary operators. It can 
distribute the computation as well as 
the memory and disk requirements across different nodes.
When symmetry is used, the matrix representation of operators becomes
block-sparse. Based on this, Kurashige and Yanai\cite{kurashige2009high} 
developed a parallelization strategy by distributing symmetry sectors.
Li and Chan\cite{li2017spin} realized the parallelization based on 
the sum of matrix product operators (MPO) representation\cite{chan2016matrix}
for the Hamiltonian in the context of spin-projected DMRG. 
Brabec et al.\cite{brabec2021massively} combined operator and symmetry sector parallelisms
and applied the resulting program to the FeMoco cluster 
(with an active space comprising 113 electrons in 76 orbitals\cite{li2019electronic2})
with a bond dimension equal to 6000 using approximately 2000 CPU cores.
Zhai et al.\cite{zhai2021low} combined the real-space\cite{stoudenmire2013real}
and the sum of MPO\cite{chan2016matrix} parallelisms and achieved a good parallel scaling
up to thousands of CPU cores. The development of heterogeneous parallel strategy for \emph{ab initio} DMRG appears only very recently. Menczer and Legeza\cite{menczer2023massively,menczer2023boosting} reported a single-node multi-GPU parallelization of \emph{ab initio} DMRG and applied to
the FeMoco with $D=4096$\cite{menczer2023boosting}. 
Apart from these parallelization strategies developed
for \emph{ab initio} DMRG, other parallelization strategies 
have also been introduced in the context of DMRG for model Hamiltonians\cite{hager2004parallelization,
levy2020distributed,nemes2014density,elwasif2018miniapp,li2020numerical,chen2020improved,chen2021real,ganahl2023density}. A closely related work to the present study
is the application of batched GEMM on GPU using the MAGMA\cite{abdelfattah2016performance} 
library in the DMRG++ code\cite{elwasif2018miniapp}. 
For a review of previous parallel DMRG studies, see Ref. \cite{tian2023high}. 
To the best of our knowledge, distributed multi-GPU parallelization of \emph{ab initio} DMRG has not been reported.

In this work, we present a distributed multi-GPU \emph{ab initio} DMRG algorithm. The central idea is to perform the most computationally intensive part, that is, the multiplication of $O(K^2)$ operators with a trial wavefunction, by combining operator parallelism for distributing
the workload with a batched algorithm for performing contractions on GPU. 
The remainder of the paper is organized as follows. In Sec.
\ref{sec:theory}, we recapitulate DMRG and introduce our distributed multi-GPU parallelization.
In Sec. \ref{sec:results}, we benchmark the algorithm by applying it to the ground-state energy of an active space model (114 electrons in 73 active orbitals) of the P-cluster\cite{li2019electronic} and analyze the performance of our implementation. The capability of the present algorithm
is demonstrated by reaching an unprecedentedly large bond dimension
$D=14000$ for the P-cluster model, which is nearly three times larger than the bond dimensions reported in previous DMRG calculations for the same system using only CPUs. Conclusions and outlooks for future directions are presented in Sec. \ref{sec:conclusion}.

\section{Theory and algorithm}\label{sec:theory}
\subsection{Recapitulation of DMRG}
The DMRG algorithm is a variational algorithm aiming to find the low-lying eigenstate of a Hamiltonian
\begin{eqnarray}
\hat{H}\Psi = E\Psi,\label{eq:SE}
\end{eqnarray}
with
\begin{eqnarray}
\hat{H} = \sum_{pq}h_{pq}\hat{a}_p^\dagger \hat{a}_q + \frac{1}{4}\sum_{pqrs}
\langle pq\|rs\rangle \hat{a}_p^\dagger \hat{a}_q^\dagger \hat{a}_s \hat{a}_r,\label{eq:Hsq}
\end{eqnarray}
by a special low-rank approximation, called matrix product states\cite{schollwock2011density,chan2016matrix} (MPS), where the many-body wavefunction $\Psi$ 
in the Fock space is approximated as
\begin{eqnarray}
\Psi^{n_1n_2\cdots n_K} \approx \sum_{\{\alpha_k\}} A^{n_1}_{\alpha_1}[1]
A^{n_2}_{\alpha_1\alpha_2}[2]\cdots 
A^{n_k}_{\alpha_{k-1}\alpha_k}[k]\cdots
A^{n_K}_{\alpha_{K-1}}[K],\label{eq:MPS}
\end{eqnarray}
with $n_k\in\{0,1,2,3\}$ for the occupation patterns
$\{|vac\rangle,|k_\beta\rangle,|k_\alpha\rangle,|k_\alpha k_\beta\rangle\}$.
Here, the tensors at the boundary $A^{n_1}_{\alpha_1}[1]$ and $A^{n_K}_{\alpha_{K-1}}[K]$ are matrices, while those $A^{n_k}_{\alpha_{k-1}\alpha_k}[k]$ in the middle
of the MPS chain are rank-3 tensors.
In practice, a maximal value $D$ for the virtual index $\alpha_k$ 
is given as an input for the DMRG algorithm. It is usually referred as the bond dimension,
and controls the accuracy of the DMRG calculations.
The set of tensors $\{A^{n_k}_{\alpha_{k-1}\alpha_k}[k]\}$
is to be found by minimizing the energy function
\begin{eqnarray}
E[\{A^{n_k}_{\alpha_{k-1}\alpha_k}[k]\}]=
\langle \Psi(\{A^{n_k}_{\alpha_{k-1}\alpha_k}[k]\})|
\hat{H}|\Psi(\{A^{n_k}_{\alpha_{k-1}\alpha_k}[k]\})\rangle,
\end{eqnarray}
subject to the normalization condition $
\langle \Psi(\{A^{n_k}_{\alpha_{k-1}\alpha_k}[k]\})|\Psi(\{A^{n_k}_{\alpha_{k-1}\alpha_k}[k]\})\rangle=1$.
Since the number of parameters in each tensor
$A^{n_k}_{\alpha_{k-1}\alpha_k}[k]$ scales as $O(D^2)$,
the total of parameters in an MPS is of $O(KD^2)$. In DMRG,
the optimization is carried out in a sequential way (called \emph{sweep}),
in which all other tensors are kept fixed when a one-site tensor $C^{n_k}_{\alpha_{k-1}\alpha_{k}}[k]$ or a two-site tensor $C^{n_k n_{k+1}}_{\alpha_{k-1}\alpha_{k+1}}[k,k+1]\triangleq (A[k]A[k+1])^{n_k n_{k+1}}_{\alpha_{k-1}\alpha_{k+1}}$ is being optimized in one-dot or two-dot sweep algorithms.

In the commonly used two-dot algorithm, Eq. \eqref{eq:SE} is solved in a direct product space
$V_L\otimes V_{C_1}\otimes V_{C_2}\otimes V_R=
\{|l_{k-1}n_{k}n_{k+1}r_{k+1}\rangle\}$,
which is a subspace of the entire Fock space defined by the
contracted basis $|l_{k-1}n_{k}n_{k+1}r_{k+1}\rangle$,
using an iterative algorithm
such as the Davidson algorithm\cite{davidson1975theiterative}. In this case,
the Hamiltonian $\hat{H}$ \eqref{eq:Hsq} can be written as a sum of product form
\begin{eqnarray}
\hat{H} = \sum_{\mu} \hat{O}^L_\mu \hat{O}^{C_1}_\mu\hat{O}^{C_2}_\mu \hat{O}^R_\mu.\label{eq:totoalH}
\end{eqnarray}
where the number of terms ($\mu$) scales as $O(K^2)$ for the second-quantized \emph{ab initio} Hamiltonian\cite{white1999ab,chan2002highly}. The wavefunction expanded in this space reads
\begin{eqnarray}
|\Psi\rangle = \sum_{lc_1c_2r}|l c_1 c_2 r\rangle\Psi^{lc_1c_2r}.
\end{eqnarray}
In the DMRG algorithm, the Hamiltonian-wavefunction multiplication 
is usually the most time-consuming part, i.e.
\begin{eqnarray}
\sigma^{l'c_1'c_2'r'} &=&
\sum_{\mu}\sum_{lc_1c_2r}
\langle l'c_1'c_2'r'|\hat{O}^L_\mu\hat{O}^{C_1}_\mu\hat{O}^{C_2}_\mu\hat{O}^R_\mu|lc_1c_2r\rangle
\Psi^{lc_1c_2r} \nonumber\\
&=&
\sum_{\mu}\sum_{lc_1c_2r}
\zeta_{\mu lc_1c_2} [\hat{O}^L_\mu]_{l'l} [\hat{O}^{C_1}_\mu]_{c_1'c_1} [\hat{O}^{C_2}_\mu]_{c_2'c_2}
[\hat{O}^{R}_\mu]_{r'r}\Psi^{lc_1c_2 r},\label{eq:Sigma}
\end{eqnarray}
where $\zeta_{\mu lc_1c_2}$ is a factor ($\pm 1$) due to the Fermionic character of electrons.
The dimensions of $\hat{O}^{C_1}$ and $\hat{O}^{C_2}$ are 4, while the dimensions
of $\hat{O}^{L}$ and $\hat{O}^{R}$ are $D$. Therefore, the computational cost for Eq. \eqref{eq:Sigma} scales as $O(K^2D^3)$, and hence the total cost per sweep is $O(K^3D^3)$.
After a new wavefunction $\Psi^{lc_1c_2r}$ is obtained in the Davidson step,
the site tensor $A[k]$ can be updated using singular value decomposition (SVD)
or diagonalizing the reduced density matrix\cite{white1992density}. 
This is referred to as the decimation step. With the new $A[k]$, a renormalization
step is performed to produce the updated $\hat{O}^L$ or $\hat{O}^R$ for the next
cycle, e.g.
\begin{eqnarray}
[\hat{O}^{\mathrm{new}}_\mu]_{r'r} = \sum_{l'c'}\sum_{lc} A_{l'r'}^{c'*}\langle l'c'|\hat{O}^{L}_\mu \hat{O}^C_{\mu}|lc\rangle A_{lr}^{c}
=\sum_{l'c'}\sum_{lc} \zeta_{\mu l} A_{l'r'}^{c'*} 
[\hat{O}^L_\mu]_{l'l} [\hat{O}^{C}_\mu]_{c'c}A_{lr}^{c},\label{eq:Renorm}
\end{eqnarray}
which scales also as $O(K^3D^3)$ per sweep. For a more detailed description of the DMRG algorithm, the reader is referred to Ref. \cite{chan2002highly}.

\subsection{Distributed multi-GPU parallelization}
The distributed parallelization algorithm by Chan\cite{chan2004algorithm} distributes the computation of Eq. \eqref{eq:Sigma} by distributing $\mu$ (corresponding
to contraction pairs of normal and complementary operators\cite{chan2002highly})
to different processors at the cost of replicating
creation/annihilation operators and communications of complementary operators ($\hat{S}_p$ and $\hat{H}$) using the message passing interface (MPI)\cite{walker1996mpi}.
The leading computational cost for the Davidson and renormalization steps
becomes $O([K^3D^3+K^4D^2]/n_{p})$ with $n_{p}$ being the number of processors\cite{chan2004algorithm}. In this work, we use this algorithm to distribute operators and partition $\mu$ to different processes on difference nodes. 
Furthermore, we assume that a GPU is bound to a process, and we will use GPU
to accelerate the contractions in Eqs. \eqref{eq:Sigma} and \eqref{eq:Renorm} 
for each process. 

\begin{figure}[H]
\includegraphics[width=0.99\textwidth]{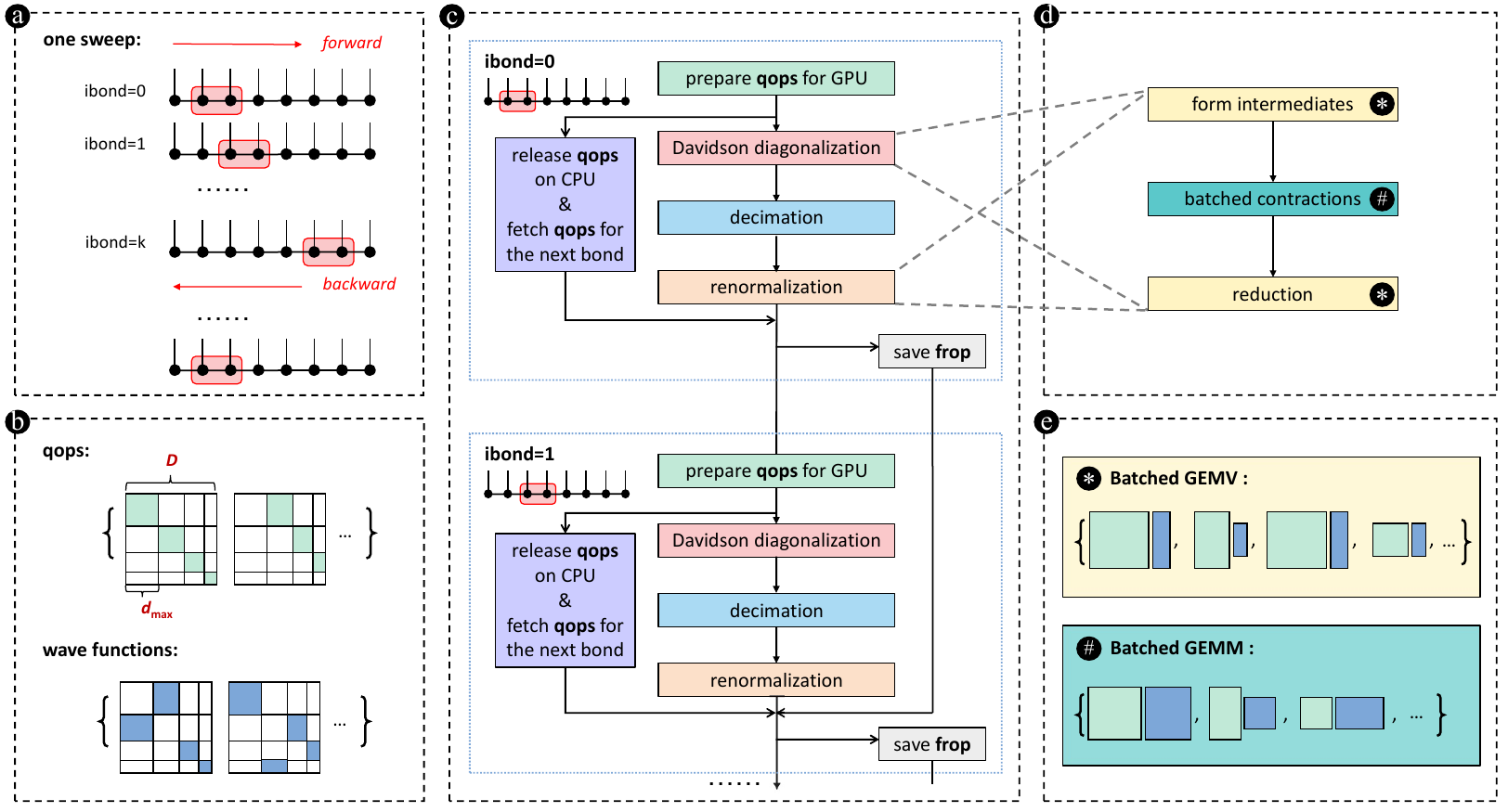}
\caption{Flowchart of the two-dot sweep algorithm implemented in this work.
Initially (ibond=0), all the necessary operators \emph{qops}, see
Eq. \eqref{eq:qops}, which are stored in a block-sparse manner, are loaded from disk to CPU memory and then copied to GPU. 
Alternatively, if NVIDIA GPUDirect Storage (GDS) is supported, then \emph{qops}
can be directly loaded to GPU.
Once the copy is finished, the CPU memory space is reused to prefetch \emph{qops} for the next bond asynchronously in order to reduce IO time.
In the Davidson diagonalization and renormalization steps, batched GEMV and GEMM are invoked on GPUs to form intermediates, perform contractions, and finally
reduce to target results. After these two steps, the newly obtained renormalized operators are copied back to CPU, and then saved to disk asynchronously. The next cycle will start with copying \emph{qops} from CPU to GPU (ibond=1). The iteration will go over the entire MPS chain and back to the initial bond, where a single DMRG sweep is finished.
}\label{fig:flowchart}
\end{figure}

In this work, we use Abelian symmetries in DMRG to reduce the computational
cost and memory consumption. Consequently, 
the operators are block-sparse and only the nonzero blocks 
are stored, see Fig. \ref{fig:flowchart}. 
We will use capital letters to represent symmetry blocks. 
In the Davidson diagonalization step, we implementation 
Eq. \eqref{eq:Sigma} via eight (instead of four as some operators
are Hermitian conjugate of basic operators) 
batches of contractions at the level of symmetry blocks, viz.
\begin{eqnarray}
I^{0/1}_{\mu L'R'C_1'C_2'LRC_1C_2}(LRC_1C_2') &=& [\hat{O}_\mu]_{C_2'C_2}\cdot\Psi(LRC_1C_2),\label{eq:sigmablk12}\\
I^{2/3}_{\mu L'R'C_1'C_2'LRC_1C_2}(LRC_1'C_2') &=& 
[\hat{O}_\mu]_{C_1'C_1}\cdot I^{0/1}_{\mu L'R'C_1'C_2'LRC_1C_2}(LRC_1C_2'),\label{eq:sigmablk34}\\
I^{4/5}_{\mu L'R'C_1'C_2'LRC_1C_2}(LR'C_1'C_2') &=& [\hat{O}_\mu]_{R'R}\cdot
I^{2/3}_{\mu L'R'C_1'C_2'LRC_1C_2}(LRC_1'C_2'),\label{eq:sigmablk56}\\
I^{6/7}_{\mu L'R'C_1'C_2'LRC_1C_2}(L'R'C_1'C_2') &=& [\hat{O}_\mu]_{L'L}\cdot
I^{4/5}_{\mu L'R'C_1'C_2'LRC_1C_2}(LR'C_1'C_2'),\label{eq:sigmablk78}\\
\sigma(L'R'C_1'C_2') &=& 
\sum_{\mu}\sum_{LRC_1C_2} \zeta_{\mu LC_1C_2} \cdot I^{6/7}_{\mu L'R'C_1'C_2'LRC_1C_2}(L'R'C_1'C_2'),\label{eq:sigmablk}
\end{eqnarray}
where $\sigma(L'R'C_1'C_2')$ represents the symmetry block of $\sigma^{lc_1c_2r}$ \eqref{eq:Sigma} 
stored in C order, and the subscript in $I_{\mu L'R'C_1'C_2'LRC_1C_2}$ indicates that
the intermediate is obtained from the term $\mu$ and the combination of
symmetry blocks $L'R'C_1'C_2'$ and $LRC_1C_2$.
For nonrelativistic or spin-free scalar relativistic Hamiltonians 
with particle number and spin projection symmetry, $[\hat{O}_\mu]_{C_1'C_1}$
and $[\hat{O}_\mu]_{C_2'C_2}$ are just numbers, and hence Eqs. \eqref{eq:sigmablk12}
and \eqref{eq:sigmablk34} can be omitted by absorbing these
factors into the scalar factor in the final step \eqref{eq:sigmablk}.
For relativistic Hamiltonian including spin-orbit couplings, the
spin projection is no longer a good quantum number, and
$[\hat{O}_\mu]_{C_1'C_1}$ and $[\hat{O}_\mu]_{C_2'C_2}$ 
can be either a scalar or a two-by-two matrix.
In this case, all the eight batches of contractions are needed.
Each batch of contractions in Eqs. \eqref{eq:sigmablk12}-\eqref{eq:sigmablk78}
can be recast into a batch of independent 
dense matrix-matrix multiplications (see Fig. \ref{fig:flowchart}).
The preprocessing of different terms in the Hamiltonian ($\mu$), 
symmetry blocks (i.e., $L$, $R$, etc.), and the expansion
of Eqs. \eqref{eq:sigmablk12}-\eqref{eq:sigmablk78} into
dense matrix-matrix multiplications are
carried out on CPU. Each batch of matrix multiplications takes the form 
\begin{eqnarray}
\{\mathbf{C}_1=\mathbf{A}_1\mathbf{B}_1, \; \cdots,\;
\mathbf{C}_T=\mathbf{A}_T\mathbf{B}_T\}.\label{eq:batchGEMM}
\end{eqnarray}
However, the dimensions of
the matrices are not the same, and its efficient computation 
using batched GEMM on GPU will be discussed in details in the next section.
A similar strategy is developed for the renormalization step \eqref{eq:Renorm}, viz.
\begin{eqnarray}
I_{\mu L'R'C'LRC}^{0/1}(LRC') &=& [\hat{O}^{C}_\mu]_{C'C}\cdot A(LRC), \label{eq:renorm12}\\
I_{\mu L'R'C'LRC}^{2/3}(L'RC') &=& [\hat{O}^{L}_\mu]_{L'L}\cdot I_{\mu L'R'C'LRC}^{0/1}(LRC'), \label{eq:renorm34}\\
I_{\mu L'R'C'LRC}^{4}(R'R) &=& A^{*}(L'R'C')\cdot I_{\mu L'R'C'LRC}^{2/3}(L'RC'), \label{eq:renorm56}\\
\protect[\hat{O}^{\mathrm{new}}_\mu]_{R'R} &=& \sum_{L'C'LC} \zeta_{\mu L} \cdot I_{\mu L'R'C'LRC}^{4}(R'R).
\label{eq:renormFinal}
\end{eqnarray}

In summary, we realize the contractions in both the Davidson and renormalization steps
using batched linear algebra on GPU via three steps (see Fig. \ref{fig:flowchart}). First, 
in order to reduce the memory cost, the necessary blocks of $\hat{O}^L$ or $\hat{O}^R$ are formed from normal/complementary operators (e.g., $\hat{O}^R=\sum_{s}\alpha_s \hat{a}_s^R$)
on the fly using batched matrix-vector multiplications (GEMV), which we will
refer as the step for intermediates. Second, batched GEMM are applied for Eqs. \eqref{eq:sigmablk12}-\eqref{eq:sigmablk78} or Eqs. \eqref{eq:renorm12}-\eqref{eq:renorm56},
which will be referred as the GEMM step for simplicity. The final step for Eq. \eqref{eq:sigmablk} or Eq. \eqref{eq:renormFinal} referred as the reduction
is achieved via batched GEMV for different combinations
of operators ($\mu$) and symmetry blocks.

\subsection{Implementation details}\label{sec:implementation}
This distributed multi-GPU \emph{ab initio} DMRG algorithm is implemented in our in-house
program \textsc{Focus}\cite{li2021expressibility,li2023time} written
in C++ and CUDA. Here we provide a more detailed description of the implementation of batched contractions as well as other optimizations. The flowchart of the two-dot sweep algorithm implemented in this work is shown in Fig. \ref{fig:flowchart}. The detailed explanations are as follows. We refer the coordinate pair $(k,k+1)$ in the MPS chain \eqref{eq:MPS} as a \emph{bond}. A sweep consists of a sequence
of bonds to be iterated, where at each bond either $A[k]$ or $A[k+1]$ will be updated
depending on the sweep direction is forward or backward (see Fig. \ref{fig:flowchart}a).
All the necessary normal and complementary operators\cite{xiang1996density,white1999ab,chan2002highly} for forming $\hat{O}^L$, $\hat{O}^R$, $\hat{O}^{C_1}$, and $\hat{O}^{C_2}$ in Eq. \eqref{eq:totoalH} at a given bond are named \emph{qops}, which include
\begin{eqnarray}
\{\hat{a}_p^{X\dagger},\hat{A}^X_{pq},\hat{B}^X_{ps},\hat{P}^X_{pq},\hat{Q}^X_{ps},\hat{S}^X_{p},\hat{H}^X\},\quad X\in\{L,R,C_1,C_2\},\label{eq:qops}
\end{eqnarray}
where
\begin{eqnarray}
\hat{A}^X_{pq} &=& \hat{a}_{p}^{X\dagger} \hat{a}_{q}^{X\dagger}, \\
\hat{B}^X_{ps} &=& \hat{a}_{p}^{X\dagger} \hat{a}^{X}_{s}, \\
\hat{P}^X_{pq} &=& \sum_{s<r}\langle pq\|sr\rangle \hat{a}_{r}^X \hat{a}_{s}^X, \\
\hat{Q}^X_{ps} &=& \sum_{qr}\langle pq\|sr\rangle \hat{a}_{q}^{X\dagger} \hat{a}_{r}^X, \\
\hat{S}_{p} &=& \sum_q\frac{1}{2}h_{pq}\hat{a}_{q}^X + \sum_{q,s<r}\langle pq\|sr\rangle \hat{a}_{q}^{X\dagger} \hat{a}_{r}^X \hat{a}_{s}^X.
\end{eqnarray}
These operators are stored in a block-sparse manner (see Fig. \ref{fig:flowchart}b) using Abelian symmetries. At the initial bond, we first load
\emph{qops} from disk to CPU and then copy them from CPU to GPU (see Fig. \ref{fig:flowchart}c). Note that 
if NVIDIA GPUDirect Storage (GDS) is supported,
\emph{qops} can be directly loaded to GPU.
In our implementation, we assume that both the CPU and GPU memories are large enough to hold the \emph{qops} at a given bond, otherwise more nodes are needed. This will eventually become the bottleneck for extremely large-scale calculations, because the memory requirement for one process
scales as $O(K^2D^2/n_{p}+KD^2)$, where $O(KD^2)$ comes
from the storage for $\hat{a}_p$ and $\hat{S}_p$, which 
are duplicated in each process\cite{chan2002highly}
and will eventually become the bottleneck.
Therefore, further optimizations are necessary to overcome this limitation in future. Once the copy is finished, we release
the \emph{qops} for the current bond on CPU, and use the memory space to prefetch
\emph{qops} for the next bond asynchronously.
In this way, loading \emph{qops} from disk to CPU memory,
which will become time consuming for large-scale calculations,
can be hided by the the Davidson diagonalization and renormalization 
steps. Similarly, after the renormalized operators are formed on GPU, 
they are copied back to CPU memory and then stored on disk asynchronously, which further hides the cost for IO by overlapping the saving process with the iteration for the next bond.

\begin{figure}[H]
    \centering
    \begin{tabular}{ccc}
    \includegraphics[width=0.31\textwidth]{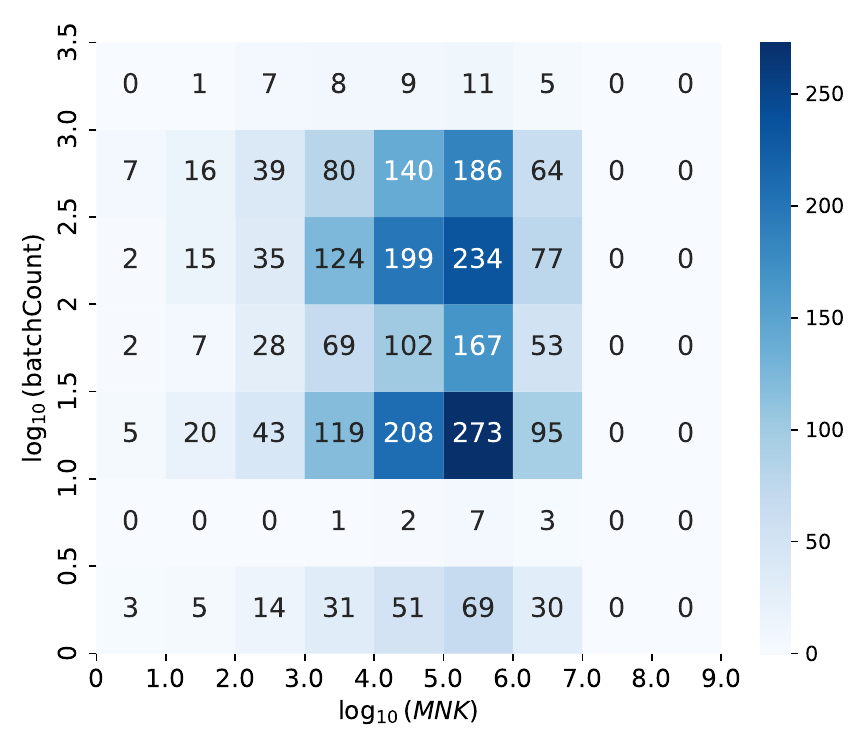} & 
    \includegraphics[width=0.31\textwidth]{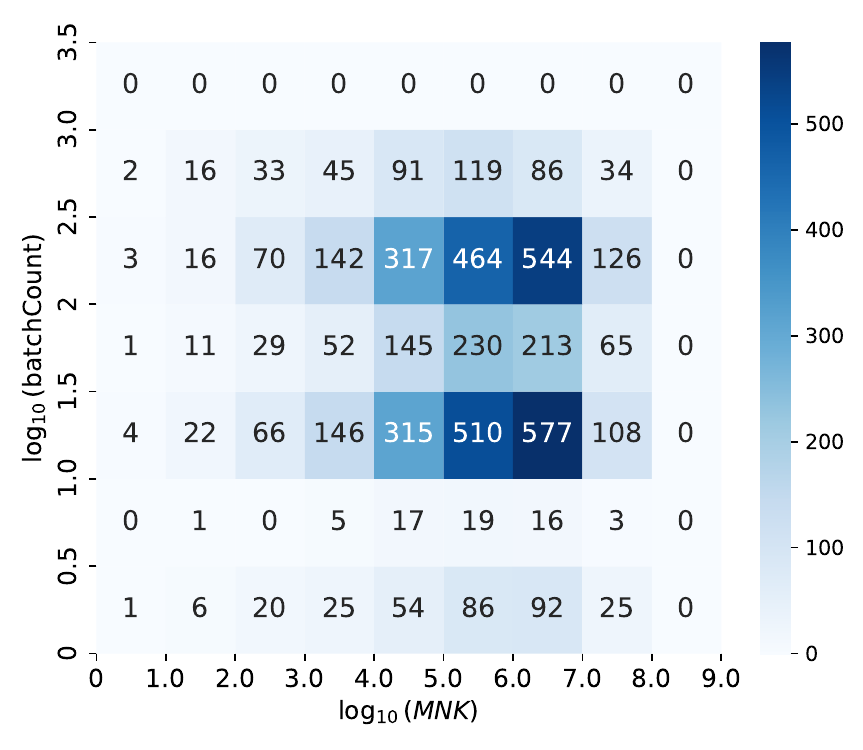} &
    \includegraphics[width=0.31\textwidth]{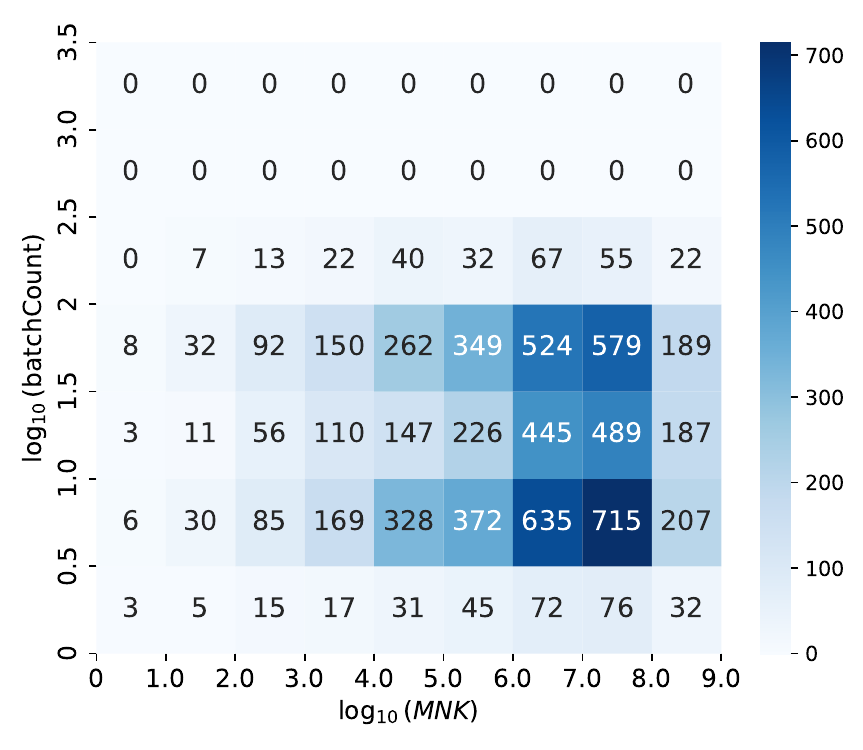} 
\\
    (a) $D=1000$ and $n_p=4$  & (c) $D=2000$ and $n_p=4$ & 
    (e) $D=5000$ and $n_p=16$ \\
    \includegraphics[width=0.31\textwidth]{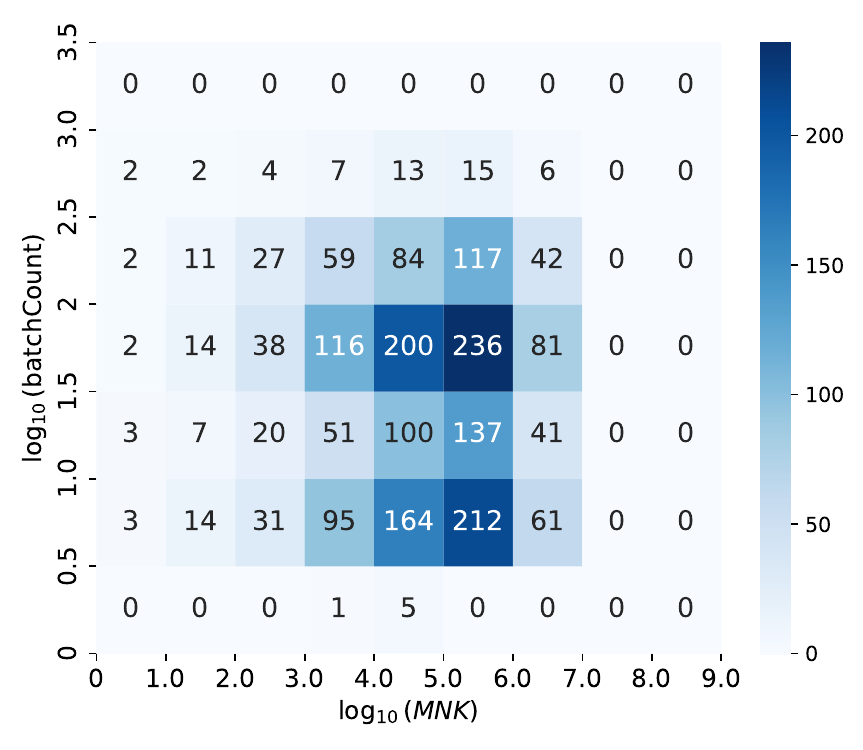} &
    \includegraphics[width=0.31\textwidth]{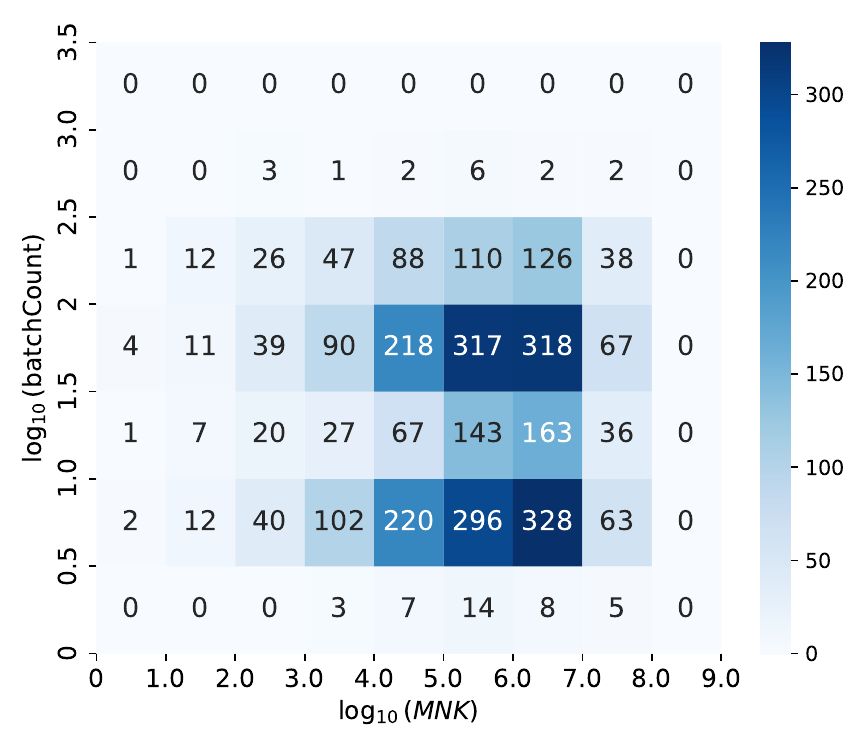}    
& 
	\includegraphics[width=0.31\textwidth]{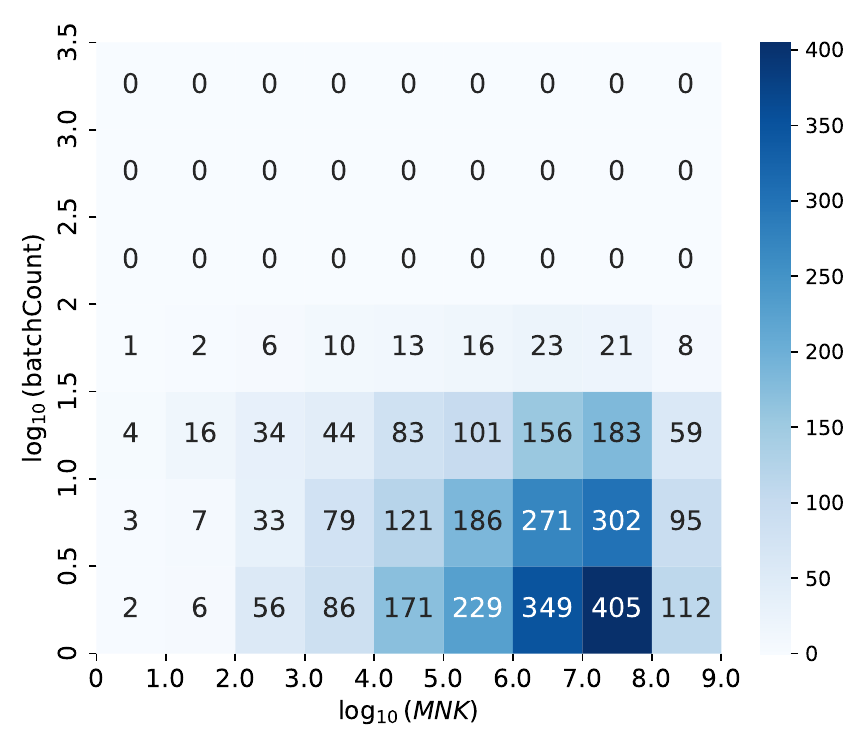} \\
    (b) $D=1000$ and $n_p=16$ & (d) $D=2000$ and $n_p=16$ 
 & (f) $D=5000$ and $n_p=128$ \\
    \end{tabular}
    \caption{Distribution
of groups for the batched GEMM in Eq. \eqref{eq:sigmablk56}
appeared in the DMRG calculations of the P-cluster model (ibond=34)
with different bond dimensions $D$ and number of processes $n_{p}$. The data measured 
at rank 0 are displayed. The $x$ axis represents the cost of a single GEMM
$C_{mn}=\sum_{k=1}^{K}A_{mk}B_{kn}$ ($m=1,\cdots, M$ and $n=1,\cdots,N$),
while the $y$ axis represents the number of GEMMs of the same sizes ($M$, $N$, and $K$),
which is referred as batchCount. The number shown in the graph represents 
the number of batched GEMM, that is, the number of function calls of the cuBLAS batched GEMM kernel, within the intervals.}\label{fig:batch}
\end{figure}

In both the Davidson and renormalization steps, batched GEMM of form
Eq. \eqref{eq:batchGEMM} is the most computationally intensive part. 
In the DMRG++ code\cite{elwasif2018miniapp} for model Hamiltonians,
it is handled with the MAGMA library\cite{abdelfattah2016performance} (\texttt{magmablas\_dgemm\_vbatched}).
In our test, we found the performance of such implementation for \emph{ab initio} DMRG
typically gives 5-6 TFLOPS on NVIDIA A100 GPU, which is about half 
of the peak performance for double precision using CUDA cores (9.7 TFLOPS).
A more severe problem is that Tensor Cores are not supported yet by \texttt{magmablas\_dgemm\_vbatched}.
To achieve better performance, a key observation is that different from
the case for model Hamiltonians, for \emph{ab initio} Hamiltonians \eqref{eq:totoalH}
there are many operators in Eqs. \eqref{eq:Sigma} and \eqref{eq:Renorm} 
of the same kind (e.g. $\hat{a}_p^{X\dagger}$
for different $p$, $\hat{A}_{pq}^X$ for different $p<q$, etc.).
This implies that GEMM encountered in Eq. \eqref{eq:batchGEMM}
can be classified into groups by a simple sort, where within each group the dimensions are the same.
To illustrate this fact, Figure \ref{fig:batch} displays the distribution
of groups for the batched GEMM in Eq. \eqref{eq:sigmablk56}
appeared in the calculations of
the P-cluster model with different bond dimensions $D$
and different number of processes $n_{p}$.
In Table \ref{tab:stat}, we show the statistical analysis
for the obtained results in Fig. \ref{fig:batch}.
It is seen that the grouping strategy is very effective, 
in particular, for small $n_{p}$. We will refer
the number of GEMMs of the same sizes as the batchCount
as used in cuBLAS (\texttt{cublasDgemmBatched}).
As shown in Table \ref{tab:stat}, for $n_p=4$ with $D=1000$ and $D=2000$, the averaged batchCount $N_1/N_2$ are greater than 100.
This allows to effectively use the highly optimized kernel
in cuBLAS (\texttt{cublasDgemmBatched}), which currently only
supports batched GEMM for matrices with the same size
and can use Tensor Cores for acceleration.
When $n_p$ is increased to 16, the averaged batchCount $N_1/N_2$ becomes
smaller, because the operators 
of the same type are eventually distributed to different processors
as $n_p$ increases. This can lower the performance for small
$D$. To take into account this, our final solution is to
combine the use of batched GEMM with CUDA streams.
For large $D$, the averaged cost for batched GEMM $C/N_2$ will become
large enough to achieve good performance even with a small averaged batchCount $N_1/N_2$, since the size of the
involved matrices will be increasingly large.
Compared with using MAGMA, this combined approach leads to a much better performance (see Sec. \ref{sec:results}).

\begin{table}[H]
\caption{
Statistical analysis for the distributions in Fig. \ref{fig:batch}.
$n_p$: number of processors; $N_1$: total number of GEMM;
$N_2$: total number of batched GEMM; $N_1/N_2$:
averaged batchCount; $C$: total cost for GEMM (or equivalently batched GEMM) given by $\sum_{i=1}^{N_1} 2M_iN_iK_i$;
$C/N_2$: averaged cost for batched GEMM.
}\label{tab:stat}
\begin{tabular}{ccccccc}
\hline\hline
$D$	&	$n_p$	&	$N_{1}$ &	$N_{2}$	&	
$N_1/N_2$ &	
$C/10^9$ &	
$C/N_2/10^6$ \\
\hline
1000	&	4	&	508073	&	2666	&	190.6	&	386.1	&	144.8	\\
1000	&	16	&	128196	&	2023	&	63.4	&	97.7	&	48.3	\\
2000	&	4	&	625165	&	4972	&	125.7	&	2917.4	&	586.8	\\
2000	&	16	&	157684	&	3092	&	51.0	&	736.3	&	238.1	\\
5000	&	16	&	198853	&	6960	&	28.6	&	10323.0	&	1483.2	\\
5000	&	128	&	27256	&	3293	&	8.3	&	1408.0	&	427.6	\\
\hline\hline
\end{tabular}
\end{table}

In this work, the program was tested on two different hardware platforms shown in Table \ref{tab:platforms}. In the Davidson diagonalization, the formation and diagonalization of the Hamiltonian in a subspace, the computation of residuals and preconditioning, and 
the Gram-Schmidt orthonormalization are performed on CPU
only at rank 0, while the formation of $\sigma$-vector in Eq. \eqref{eq:sigmablk} 
is performed on distributed GPUs. We use the NVIDIA Collective Communication Library (NCCL)\cite{nccl} to accelerate multi-GPU and multi-node communications 
for broadcasting trial vectors and the reduction of $\sigma$-vector in the Davidson step,
as well as the reduction of complementary operators ($\hat{S}_p$ and $\hat{H}$) 
in the renormalization step. 

Besides, we find the CPU part of the Davidson diagonalization
on platform 1 with ARM processors using 32 threads 
is much slower than that on platform 2 with Intel processor using 8 threads,
since some basic functions including \texttt{dcopy}, \texttt{dnrm2}, \texttt{dgemm} are 
slower using OpenBLAS than using MKL. To get a comparable performance, we reoptimize these
functions for ARM processors. OpenMP is used for optimizing \texttt{dcopy}, while SIMD instruction and multi-threading techniques are used for optimizing \texttt{dnrm2}. For GEMM used in
the Davidson step, we note that the special shape of matrices ($K\gg M,N$)
necessitates the adoption of thread-level parallelism for all $M$/$N$/$K$ dimensions, while OpenBLAS's inherent support for parallelism is only for the $M$ and $N$ dimensions.
Besides, we have introduced a suite of assembly-level optimizations, comprising SIMD vectorization, loop unrolling, and prefetching. These optimizations, when applied within a 32-thread configuration, yield substantial performance enhancements, resulting in impressive speedup factors ranging from 10 to 15 compared with OpenBLAS in the calculations of the P-cluster with $D=8000$.

\begin{table}[H]
\caption{Hardware platforms and software versions used in this work. The reported timing data were mostly obtained on platform 1, except for the blue lines in Fig. \ref{fig:TFLOPS}. The results in Table \ref{tab:energy}
requiring large CPU and GPU memory were obtained on platform 2.
}\label{tab:platforms}
\begin{tabular}{ccc}
\hline\hline
&	  platform 1 (ARM)      	&	  platform 2 (Intel)   	\\
\hline
CPU                                                               	&	 Huawei Kunpeng 920 	&	 Intel Xeon Gold 8358 	\\
no. of CPUs per node                                              	&	2	&	2	\\
CPU clockspeed                                       			  	&	 3.0 GHz   	&	 2.6 GHz     	\\
total no. of cores                                                	&	128	&	64	\\
CPU memory per node                                               	&	 250GB 	&	 1.5TB            	\\
GPU                                                               	&	 NVIDIA A100 40GB PCIe       	&	 NVIDIA A100 80GB SXM         	\\
no. of GPUs per node                                              	&	4	&	8	\\
performance (FP64)                                       	&	 9.7 TFLOPS     	&	 9.7 TFLOPS      	\\
performance (FP64 Tensor Core)                                       	&	 19.5 TFLOPS     	&	 19.5 TFLOPS      	\\
GPU bandwidth                                                     	&	 1555 GB/s       	&	 2039 GB/s        	\\
CUDA version                                                      	&	11.4	&	11.6	\\
MPI version                                                       	&	 OpenMPI 4.1.2   	&	 OpenMPI 4.1.5    	\\
NCCL version                                                      	&	 2.16.5          	&	 2.17.1           	\\
MAGMA version                                                     	&	 2.7.1           	&	 2.7.1            	\\
BLAS/LAPACK                                                       	&	 OpenBLAS 0.3.23$^a$ 	&	 Intel MKL 2022   	\\
\hline\hline
\multicolumn{3}{l}{$^a$ Some BLAS functions (\texttt{dcopy}, \texttt{dnrm2}, and \texttt{dgemm}) are reoptimized for the ARM platform.}
\end{tabular}
\end{table}

\section{Results}\label{sec:results}
\begin{figure}[H]
\includegraphics[width=0.5\textwidth]{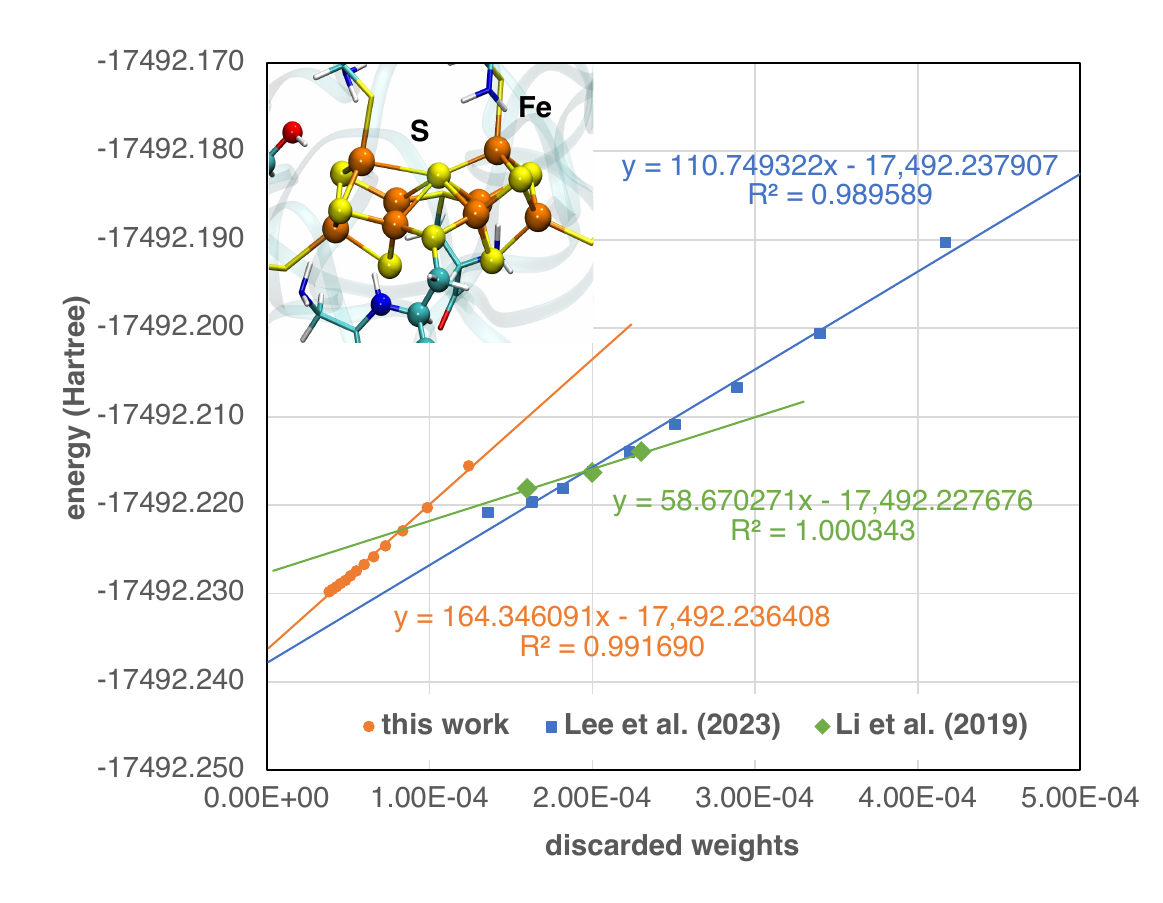}
\caption{Comparison of linear extrapolation of DMRG energies $E$ with respect to discarded weights $w$
obtained in the present and previous studies\cite{li2019electronic,lee2023evaluating}
for the P-cluster CAS(114e,73o) model.}\label{fig:extrapolation}
\end{figure}

We benchmark the performance of the above algorithm by applying it to the ground-state problem of an active space model (114 electrons in 73 active orbitals) defined in Ref. \cite{li2019electronic} for the resting state of the P-cluster (see the inset of 
Fig. \ref{fig:extrapolation}), where the integrals and orbital orderings are available online\cite{linkToFCIDUMPpclusters}.
Due to the limitation of computational resources, we performed timing analysis using platform 1 (ARM processors) with up to 32 nodes with 128 GPUs (NVIDIA A100 40GB) and $D\le 8000$. In the last part of this section, a large-scale calculation using platform 2 (Intel processors),
which has larger CPU and GPU memories per node, 
was reported using all the 6 nodes available with 48 GPUs (NVIDIA A100 80GB) in total and $D$ up to 14000.

\begin{figure}[H]
\begin{tabular}{cc}
\includegraphics[width=0.48\textwidth]{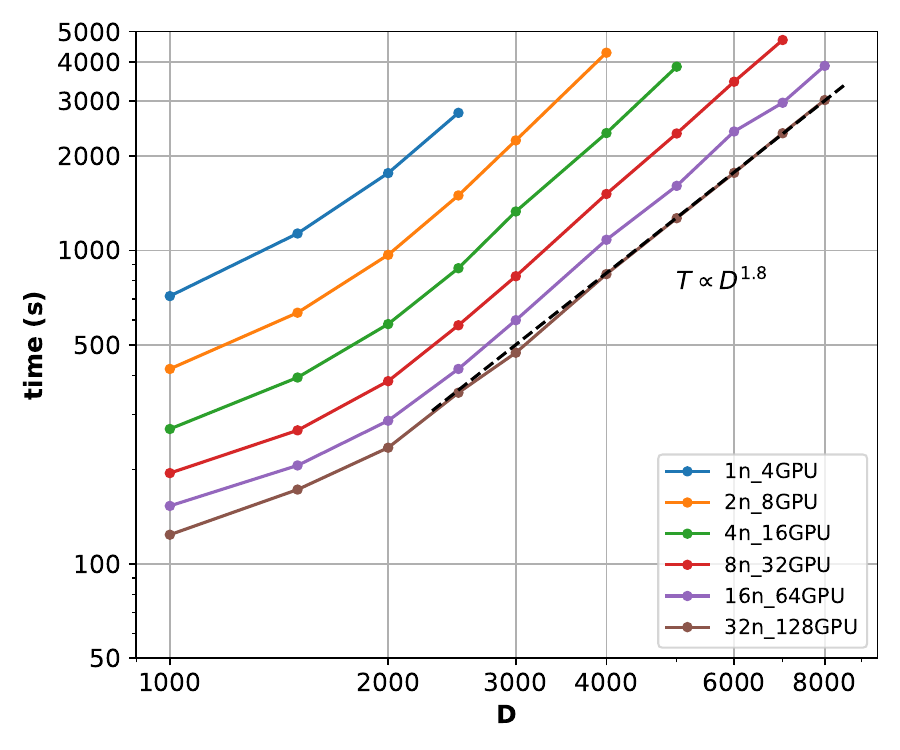} &
\includegraphics[width=0.48\textwidth]{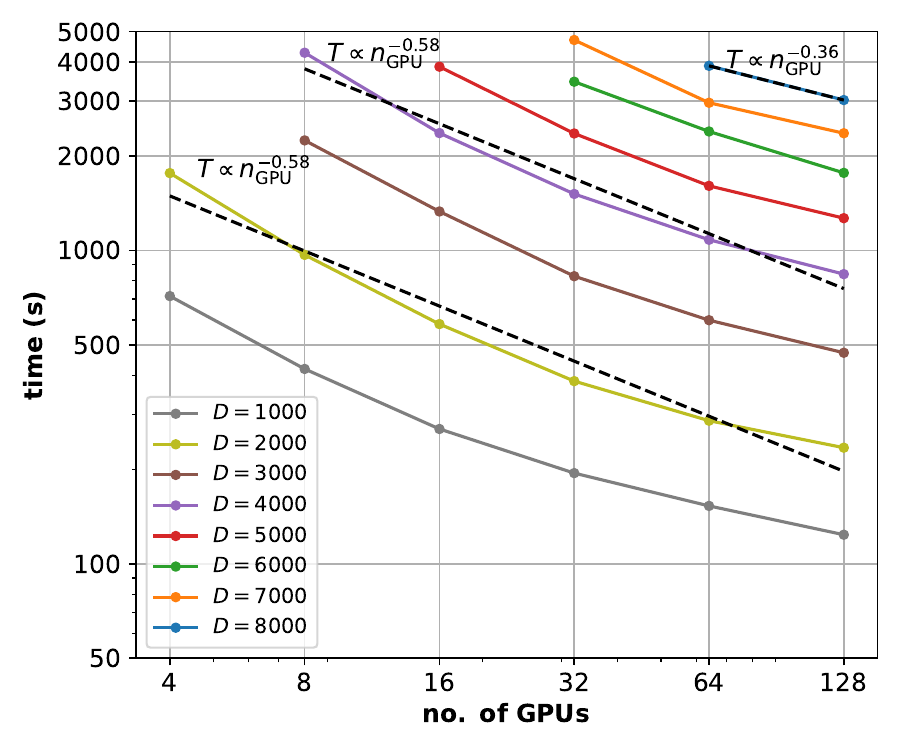} \\
\multicolumn{2}{c}{(a) Wall time for performing a single sweep $T_{\mathrm{sweep}}$ as a function of $D$ or $n_{\rGPU}$.} \\
\includegraphics[width=0.48\textwidth]{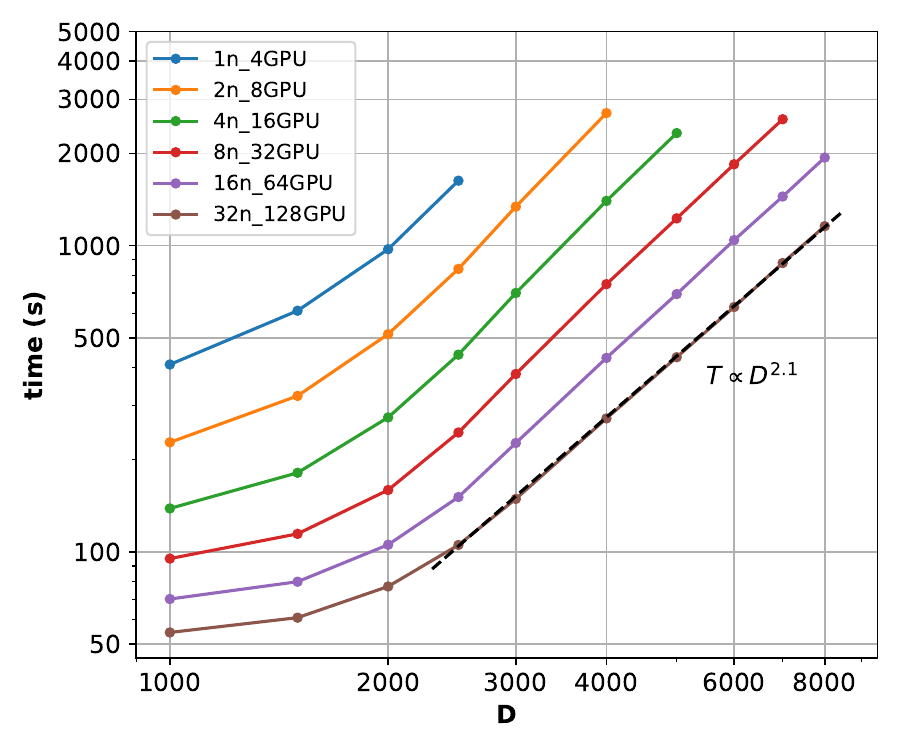} &
\includegraphics[width=0.48\textwidth]{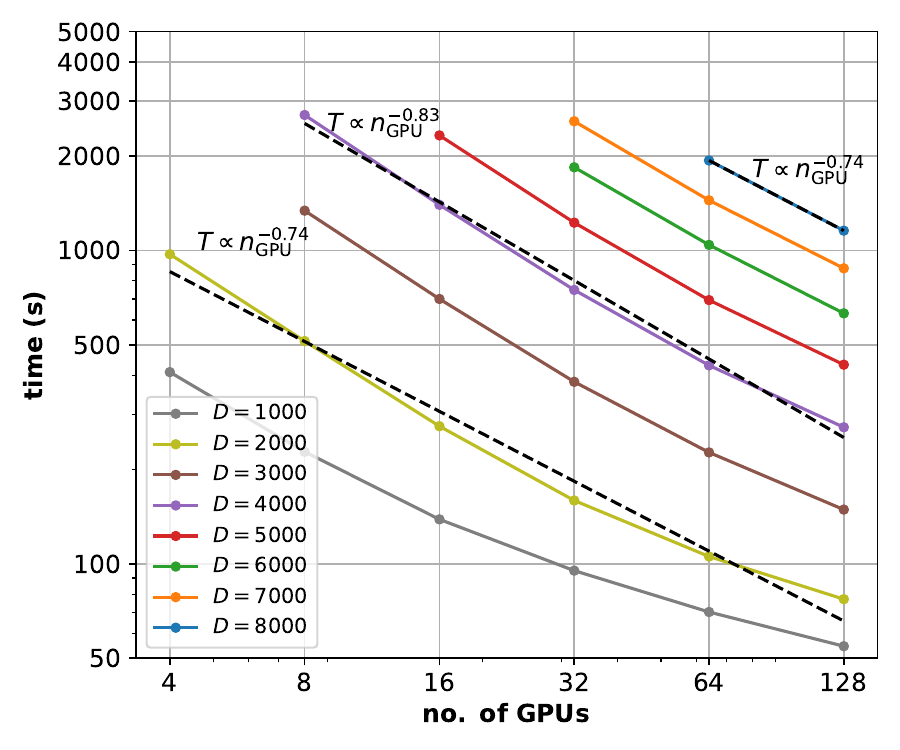} \\
\multicolumn{2}{c}{(b) Wall time for performing Eq. \eqref{eq:Sigma} in the Davidson step 
$T_{\mathrm{Hx}}$ as a function of $D$ or $n_{\rGPU}$.} \\
\end{tabular}
\caption{Wall time of performing a single sweep $T_{\mathrm{sweep}}$ and that for performing Eq. \eqref{eq:Sigma} in the Davidson step of a single sweep $T_{\mathrm{Hx}}$
as a function of the bond dimension $D$ or the number of GPUs $n_{\rGPU}$
obtained on platform 1.}
\label{fig:time}
\end{figure}

Figure \ref{fig:time} shows the wall time for performing a single sweep and that for performing Eq. \eqref{eq:Sigma} in the Davidson step (referred as Hx for simplicity in the later context) of a single sweep, denoted by $T_{\mathrm{sweep}}$ and $T_{\mathrm{Hx}}$, respectively,
as a function of the bond dimension $D$ or the number of GPUs $n_{\rGPU}$.
We find that both $T_{\mathrm{sweep}}$ and $T_{\mathrm{Hx}}$ scale roughly quadratically for large $D$, which is lower than the formal scaling $O(D^3)$, while below $D=2500$ 
the scaling is lower than quadratic. These behaviors are attributed to the use
of symmetries and GPUs. In Table \ref{tab:speedup}, we list
the maximal dimension of symmetry sectors $d_{\max}$
at a given bond dimension $D$ measured in the middle of the MPS chain.
We find $d_{\max}$ and $D$ obey a simple linear relation for this system
\begin{eqnarray}
d_{\max}(D)=0.128D+51.8.\label{eq:dmax}
\end{eqnarray}
Therefore, compared with the case without symmetry ($d_{\max}=D$),
using Abelian symmetry leads to a 7.8 fold 
reduction of $d_{\max}$. This results in a very high 
block-sparsity in the operators and wavefunctions.
Thus, the lowering from cubic to quadratic scaling is mainly due to the
use of symmetry. Ref. \cite{zhai2023block2} reported
a similar scaling with respect to $D$, $O(D^{2.40})$ and
$O(D^{2.18})$ for non-spin-adapted and spin-adapted DMRG, respectively,
using 24 CPU cores for iron sulfur dimer. However,
the gradual transition to quadratic scaling is not observed for small $D$.
We attribute such transition to the use of GPU, because below certain $d_{\max}$
the workload is not enough to fill the GPU.

\begin{table}[H]
\caption{Speedup for $T_{\mathrm{sweep}}$ and $T_{\mathrm{Hx}}$ (in parenthesis) 
for different bond dimensions $D$ using different number of nodes on platform 1,
where each node has 4 GPUs. The column $d_{\max}$ lists the maximal dimension of 
symmetry sectors at a given bond dimension $D$ measured in the middle of the MPS chain (ibond=34).
The last column shows the computed parallel efficiency. 
The symbol '-' indicates that the computation with a given $D$ cannot be 
done due to memory limitations, such that the parallel efficiency in the last column is calculated for the speedup with 32 nodes with respect to different references, where the number '1' indicates 
the reference for computing the speedup for a given $D$. 
}\label{tab:speedup}
\begin{tabular}{ccccccccc}
\hline\hline
$D$	    &	$d_{\max}$ & 1n	&	2n		&	4n		&	8n		&	16n		&	32n		&	efficiency	\\
\hline
1000	&   167    &	1	&	1.7	(1.8)	&	2.6	(2.9)	&	3.6	(4.3)	&	4.5	(5.9)	&	5.6	(7.5)	&	0.17	(0.24)	\\
2000	&	311    &    1	&	1.8	(1.9)	&	3.0	(3.5)	&	4.6	(6.1)	&	6.2	(9.2)	&	7.5	(12.6)	&	0.23	(0.40)	\\
3000	&	442    &    -	&	1        	&	1.7	(1.9)	&	2.6	(3.6)	&	3.6	(6.0)	&	4.5	(9.0)	&	0.28	(0.56)	\\
4000	&	571    &    -	&	1		    &	1.8	(1.9)	&	2.8	(3.6)	&	4.0	(6.3)	&	5.1	(9.9)	&	0.32	(0.62)	\\
5000	&	697    &   -	&	-		    &	1		    &	1.6	(1.9)	&	2.4	(3.4)	&	3.0	(5.4)	&	0.38	(0.67)	\\
6000	&	820    &   -	&	-		    &	-		    &	1		    &	1.4	(1.8)	&	2.0	(2.9)	&	0.49	(0.73)	\\
7000	&	944    &   -	&	-		    &	-		    &	1		    &	1.6	(1.8)	&	2.0	(2.9)	&	0.50	(0.74)	\\
8000	&	1071   &   -	&	-		    &	-		    &	-		    &	1	        &	1.3	(1.7)	&	0.64	(0.84)	\\
\hline\hline
\end{tabular}
\end{table}

The strong scaling is also illustrated in Fig. \ref{fig:time}, and the detailed
results for speedup is shown in Table \ref{tab:speedup} for different bond dimensions
with different number of nodes. Overall, it is seen that $T_{\mathrm{Hx}}$ scales much 
better than $T_{\mathrm{sweep}}$. For instance, $T_{\mathrm{sweep}}$ for $D=8000$ with 128 GPUs becomes $2^{-0.36}=78\%$ of that with 64 GPUs, whereas $T_{\mathrm{Hx}}$ for $D=8000$ with 128 GPUs is about $2^{-0.74}=60\%$ of that with 64 GPUs. The decreases of speedup for the entire sweep is analyzed latter. Now we focus on the analysis of the speedup only for $T_{\mathrm{Hx}}$,
where a decomposition of the speedup for different parts of Hx is shown in Table \ref{tab:speedup2}.
There are several factors that affect the total speedup for Hx. Firstly,
for small bond dimensions such as $D=1000$ or $D=2000$, the speedup
for Hx is not ideal with 32 nodes, because the workload is not large
enough. This is reflected in the Fig. \ref{fig:time}(b),
where the curves start to deviate from being linear as $n_{\rGPU}$ is greater than
32 for $D=1000$ and $D=2000$. Generally, we can observe that the parallel efficiency increases 
as the bond dimension increases. Secondly, the calculation at the middle of
the MPS chain scales better than that close to the boundary of the MPS chain,
since the bond dimension close to the boundary is actually smaller than $D$.
This can be seen from the comparison in Table \ref{tab:speedup2}
between the speedups for a single sweep and the corresponding
results for the middle of the sweep.
Thirdly, the total speedup for Hx can be slightly smaller than those
for the three computational steps as shown for $D=8000$, because in addition to
computations, $T_{\mathrm{Hx}}$ also includes the time for communication between
different nodes for broadcasting the trial vector and reducing the $\sigma$-vector 
as well as the time for copying the trial vector from CPU to GPU and
the $\sigma$-vector from GPU back to CPU. These parts take less than 10\%
of $T_{\mathrm{Hx}}$ in our tests for $D\ge 5000$ with 32 nodes.

\begin{table}[H]
\caption{Speedup for different parts of $T_{\mathrm{Hx}}$ in a single sweep obtained using 32 nodes
for $D$=2000, 5000, and 8000, with respect to those obtained using 1, 4, and 16 nodes,
respectively. The three steps (intermediates, GEMM, reduction) refer to different computational
steps in performing Eqs. \eqref{eq:sigmablk12}-\eqref{eq:sigmablk},
see also Fig. \ref{fig:flowchart}d. Values in parenthesis are the corresponding speedups for the middle of the sweep (ibond=34).
}\label{tab:speedup2}
\begin{tabular}{cccccc}
\hline\hline
$D$	    &	reference & intermediates		&	GEMM		&	reduction		&	Hx		\\
\hline
2000	&	1n        & 23.4 (27.1)	&	12.2 (20.9)	&	14.1 (21.0)	&	12.6 (19.7)	\\
5000	&	4n        & 6.4  (7.4)	&	5.9 (7.1)	&	5.1  (6.5)	&	5.4 (6.6)	\\
8000	&	16n       & 1.8  (1.9)	&   1.8 (1.9)	&	1.7  (1.9)	&	1.7 (1.8)	\\
\hline\hline
\end{tabular}
\end{table}

\begin{figure}[H]
\begin{tabular}{cc}
\includegraphics[width=0.48\textwidth]{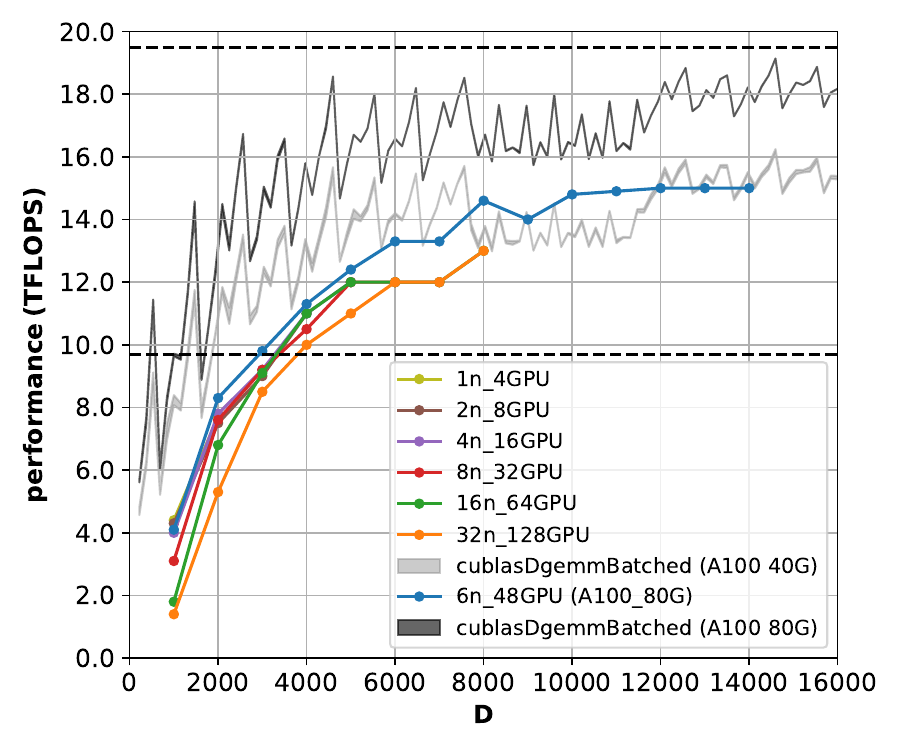} & 
\includegraphics[width=0.48\textwidth]{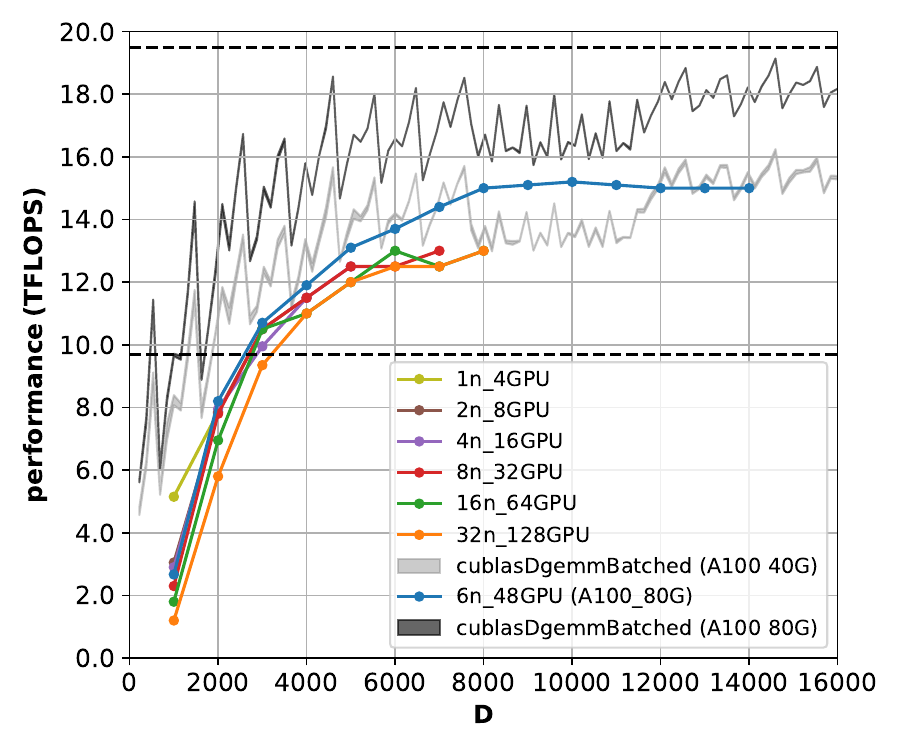} \\
\end{tabular}
\caption{Average performances for the eight batches of GEMM
in Eqs. \eqref{eq:sigmablk12}-\eqref{eq:sigmablk78} in the Davidson step
and those in Eqs. \eqref{eq:renorm12}-\eqref{eq:renorm56}
in the renormalization step as a function of $D$. The data are
measured at rank 0 in the DMRG calculation for the bond 
in the middle of the MPS chain (ibond=34). The black
dashed lines are the theoretical performances for double precision
using CUDA Cores (9.7 TFLOPS) and Tensor Cores (19.5 TFLOPS), respectively.
The black and grey lines are the measured performances of 
\texttt{cublasDgemmBatched} for a single batched GEMM
with $M=N=K=d_{\max}(D)$, where $d_{\max}(D)$ is given in Eq. \eqref{eq:dmax},
and the batchCount is determined by fixing the total memory
of all matrices ($\mathrm{batchCount}\times 3d^2_{\max}(D)$)
to be 10 GB. Except for the blue and black lines, which were obtained
on platform 2 using NVIDIA A100 SXM 80 GB, all the data were obtained 
on platform 1 using NVIDIA A100 PCIe 40 GB.
}\label{fig:TFLOPS}
\end{figure}

In Fig. \ref{fig:TFLOPS}, we show the average performances for the eight batches of GEMM
in Eqs. \eqref{eq:sigmablk12}-\eqref{eq:sigmablk78} in the Davidson step
and those in Eqs. \eqref{eq:renorm12}-\eqref{eq:renorm56}
in the renormalization step as a function of $D$ measured
at the bond in the middle of the MPS chain.
It is clear that in both cases the performances increase as $D$
increases. The performances above $D=3000$ exceed the theoretical peak performance for double-precision floating-point format (FP64) using CUDA Cores. Therefore,
using Tensor Cores leads to a significant increase in performance
and hence reduces the computational time dramatically. The peak performance
obtained on NVIDIA A100 PCIe 40 GB is about 13 TFLOPS, which is 67\% of the
peak performance for FP64 with Tensor Cores, while that obtained on NVIDIA A100 SMX 80 GB
is about 15 TFLOPS, which is 10\% higher. To better understand the performance,
we also plot the measured performances of 
\texttt{cublasDgemmBatched} for a single batched GEMM
with $M=N=K=d_{\max}(D)$, where $d_{\max}(D)$ is given in Eq. \eqref{eq:dmax},
and the batchCount is determined by fixing the total memory
of all matrices ($\mathrm{batchCount}\times 3d^2_{\max}(D)$)
to be 10 GB. This can roughly be considered as an upper bond
for our implementation. Because the actual GPU memory available
may be smaller for large $D$, and the sizes of most matrices in the batched GEMM 
are smaller than $d_{\max}$. It is seen that the performance of
our combined strategy using batched GEMM and CUDA streams is
about 85\% of this model for large $D$. Finally, we observe that
the performances with less processes are higher in particular
for $D$ less than 2000. This is because as shown in Fig. \ref{fig:batch},
the batchCount is smaller for larger $n_{\rGPU}$. This indicates
a room for further improving the performance for larger $n_{\rGPU}$ in future,
by combining different batched GEMM together in the case of small matrices.

\begin{figure}[H]
\begin{tabular}{ccc}
\includegraphics[width=0.3\textwidth]{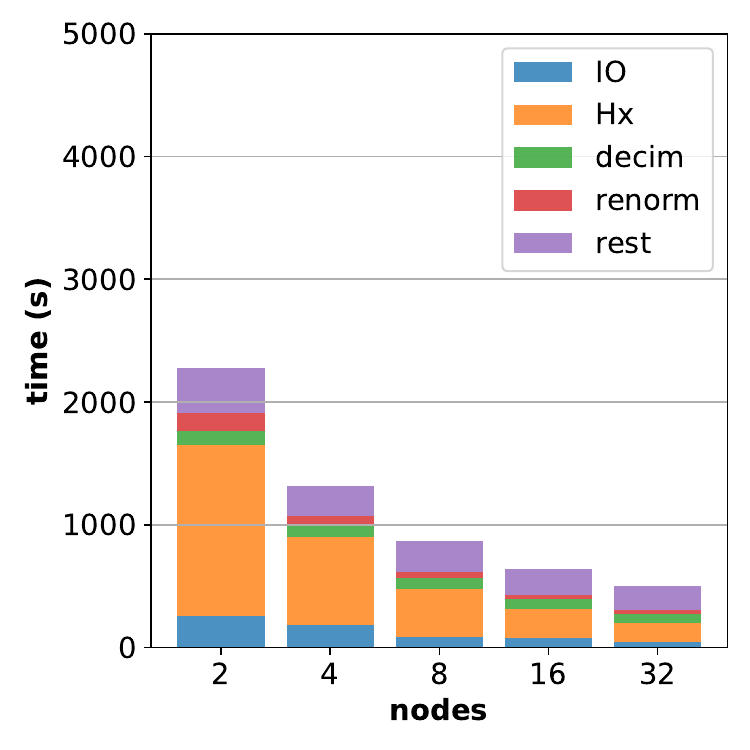} &
\includegraphics[width=0.3\textwidth]{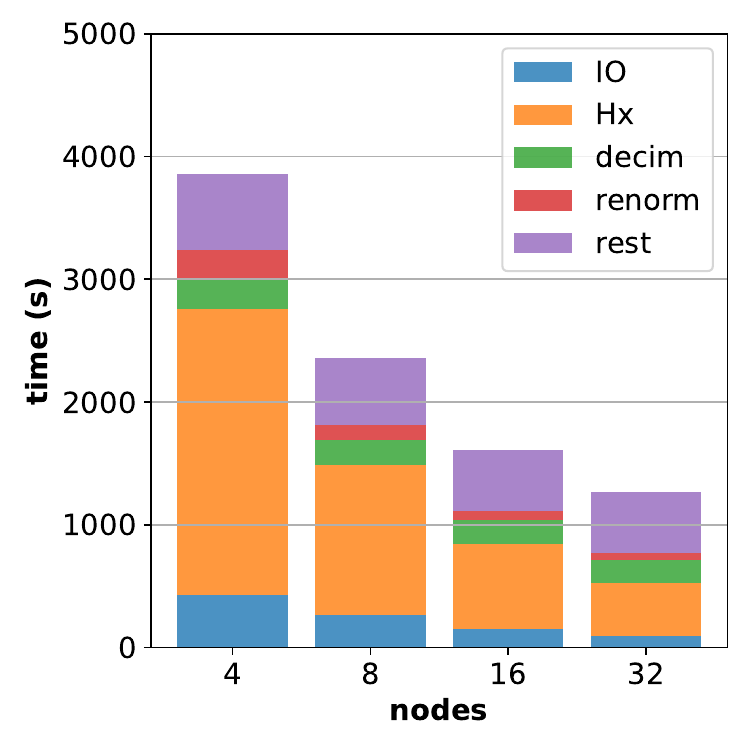} &
\includegraphics[width=0.3\textwidth]{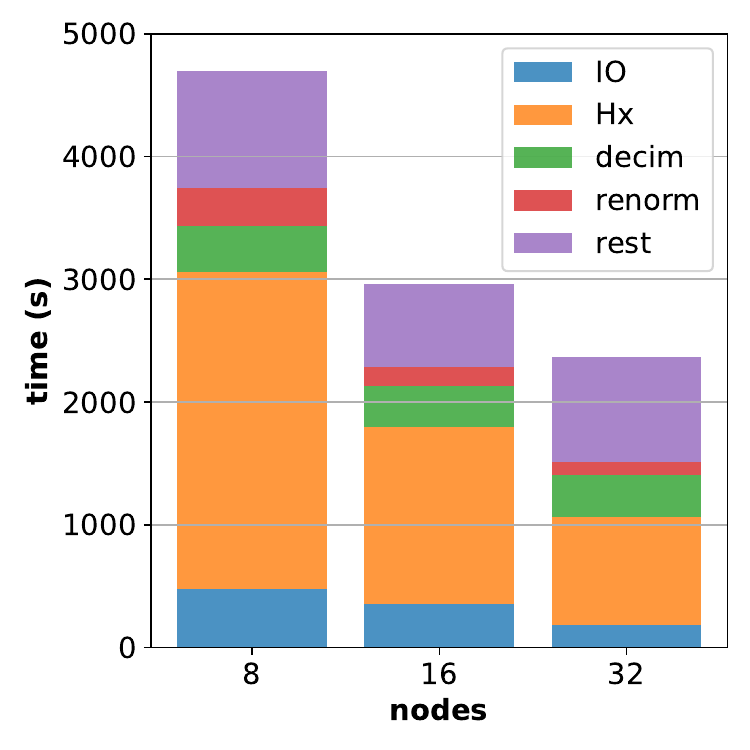} \\
(a) $D=3000$ & (b) $D=5000$ & (c) $D=7000$
\end{tabular}
\caption{Decomposition of $T_{\mathrm{sweep}}$ for $D=3000$, $D=5000$, and $D=7000$ obtained on platform 1 into IO, Hx, decimation, renormalization and the rest part, which includes the serial part of the Davidson diagonalization, 
the construction of the diagonal part of Hamiltonian, 
other communication and synchronization, etc.}\label{fig:decomposed}
\end{figure}

As mentioned before in Table \ref{tab:speedup}, the speedup for $T_{\mathrm{sweep}}$
is lower than that for $T_{\mathrm{Hx}}$. Figure \ref{fig:decomposed} shows the decomposition
of $T_{\mathrm{sweep}}$ for $D=3000$, $D=5000$, and $D=7000$ into different parts.
The part $T_{\mathrm{Hx}}$ is always the dominant part of the DMRG calculations,
which is known in previous DMRG studies\cite{chan2004algorithm,menczer2023boosting}.
In comparison, the renormalization part is quite small. 
The reduction of $T_{\mathrm{Hx}}$, $T_{\mathrm{renorm}}$, and $T_{\mathrm{IO}}$ by using more nodes is reasonably good. Thus, our goal to accelerate the most
expensive part in DMRG using multi-GPUs has been achieved. 
In the current implementation, the decimation step is simply done on
CPU, which can also be carried out on GPU using cuSOLVER. 
We note that the rest part starts to become a larger portion as the
number of nodes increases, and hence it needs to be optimized in future in order
to achieve better scalability for $T_{\mathrm{sweep}}$.
One major contribution to this part is the serial part 
in the Davidson diagonalization, which is executed only
on CPU at rank 0. If a better iterative eigensolver is implemented,
we expect that the wall time for this part can be reduced.

Finally, as an example to illustrate the capability of the present algorithm, we 
performed a large-scale DMRG calculation for the P-cluster model using platform 2
with $D$ up to 14000, and compared the obtained energies $E$
and discarded weights $w$ with those previously obtained\cite{li2019electronic,lee2023evaluating} 
using spin-adapted DMRG on CPUs in Table \ref{tab:energy}.
The largest $D$ achieved in the present work
is nearly three times larger than that
reported in previous works for the same system 
using only CPUs with MPI/OpenMP parallelization.
Calculations with larger $D$ are possible provided more GPUs are available
for meeting the memory requirement.
The energy differences between the present and
the previous works for the same $D$ are mainly due to two factors.
First, the initial guesses for MPS are different. While the initial guess in previous
studies was obtained from spin-projected MPS\cite{li2017spin} with a small bond dimension,
the initial guess in the present work is generated from the conversion
of a selected configuration interaction (SCI) wavefunction following Ref. \cite{li2021expressibility}.
Second, the present implementation only using Abelian symmetries is non-spin-adapted.
However, we found that the DMRG energy with the same $D$ obtained in the present
work is lower than the corresponding result obtained in previous works.
This is usually not the case for calculations with and without spin-adaptation, 
as spin-adaptation can lead to much lower energy for spin-coupled systems\cite{sharma2012spin}.
Therefore, the energy difference is mainly due to the different initialization strategies.
This is also reflected in the discarded weights, where the present results
are smaller than the previous results\cite{li2019electronic,lee2023evaluating} for
the same $D$.

\begin{table}[H]
\caption{Comparison of DMRG energies ($E$ in Hartree) and discarded weights ($w$) 
obtained for the P-cluster CAS(114e,73o) model. Previous results taken from
Refs. \cite{li2019electronic} and \cite{lee2023evaluating}
were obtained with spin-adapted DMRG, while the present
results were obtained without spin-adaptation.
The last line shows the extrapolated energy $E_{w=0}$ obtained via a linear 
extrapolation $E(w)=Aw+E_{w=0}$ shown in Fig. \ref{fig:extrapolation}.
}\label{tab:energy}
\centering\footnotesize
\resizebox{\columnwidth}{!}{
\begin{tabular}{ccccccccccc}
\hline\hline
\multicolumn{3}{c}{Li et al. (Ref. \cite{li2019electronic})} &&
\multicolumn{3}{c}{Lee et al. (Ref. \cite{lee2023evaluating})} &&
\multicolumn{3}{c}{this work}\\
\cline{1-3}
\cline{5-7}
\cline{9-11}
$D$	&	$w$	&	$E$	&	&	$D$	&	$w$	&	$E$	&	&	$D$	&	$w$	&	$E$	\\
\hline
3000	&	2.30$\times 10^{-4}$	&	-17492.213966	&	&	1000	&	4.17$\times 10^{-4}$	&	-17492.190311	&	&	2000	&	1.24$\times 10^{-4}$	&	-17492.215587	\\
3500	&	2.00$\times 10^{-4}$	&	-17492.216321	&	&	1500	&	3.40$\times 10^{-4}$	&	-17492.200594	&	&	3000	&	9.87$\times 10^{-5}$	&	-17492.220314	\\
4000	&	1.60$\times 10^{-4}$	&	-17492.218127	&	&	2000	&	2.89$\times 10^{-4}$	&	-17492.206725	&	&	4000	&	8.35$\times 10^{-5}$	&	-17492.222921	\\
	&		&		&	&	2500	&	2.51$\times 10^{-4}$	&	-17492.210894	&	&	5000	&	7.31$\times 10^{-5}$	&	-17492.224618	\\
	&		&		&	&	3000	&	2.23$\times 10^{-4}$	&	-17492.213953	&	&	6000	&	6.58$\times 10^{-5}$	&	-17492.225828	\\
	&		&		&	&	3500	&	2.00$\times 10^{-4}$	&	-17492.216294	&	&	7000	&	5.98$\times 10^{-5}$	&	-17492.226734	\\
	&		&		&	&	4000	&	1.82$\times 10^{-4}$	&	-17492.218146	&	&	8000	&	5.52$\times 10^{-5}$	&	-17492.227448	\\
	&		&		&	&	4500	&	1.63$\times 10^{-4}$	&	-17492.219644	&	&	9000	&	5.15$\times 10^{-5}$	&	-17492.228023	\\
	&		&		&	&	5000	&	1.36$\times 10^{-4}$	&	-17492.220847	&	&	10000	&	4.83$\times 10^{-5}$	&	-17492.228500	\\
	&		&		&	&		&		&		&	&	11000	&	4.53$\times 10^{-5}$	&	-17492.228900	\\
	&		&		&	&		&		&		&	&	12000	&	4.28$\times 10^{-5}$	&	-17492.229245	\\
	&		&		&	&		&		&		&	&	13000	&	4.04$\times 10^{-5}$	&	-17492.229539	\\
	&		&		&	&		&		&		&	&	14000	&	3.83$\times 10^{-5}$	&	-17492.229797	\\
	\\
\multicolumn{2}{l}{extrapolation}	&	-17492.227676	&	&		&		&	-17492.237907	&	&		&		&	-17492.236408	\\
\hline\hline
\end{tabular}}
\end{table}

In the last line of Table \ref{tab:energy}, we report the extrapolated
energies using a linear extrapolation with respect to the discarded weights,
see Fig. \ref{fig:extrapolation}. It is seen that 
while the earliest extrapolated energy\cite{li2019electronic} with only a few data points
seems to overestimate the ground-state energy, 
the extrapolated results obtained with the
data from Ref. \cite{lee2023evaluating} and the present work agree within 1 mH. 
Our best variational energy obtained with $D=14000$ differs from
the extrapolated energy by 6.6 mH, while the difference
for the previous results\cite{lee2023evaluating} is 17.1 mH. Thus, a reduction of the error by a factor of almost three is achieved using the new hybrid CPU-GPU implementation, and the error per metal is within 1 mH compared with the
extrapolated energies.
It deserves to point out that as shown 
in the last column of Table \ref{tab:energy}, the energy converges very slowly as $D$ increases,
which is an indication of the complex entanglement structure
for the P-cluster problem. We expect that using SU(2)
symmetry in future can accelerate the convergence.

\section{Conclusion}\label{sec:conclusion}
In this work, we presented a distributed multi-GPU \emph{ab initio} DMRG algorithm
suitable for modern heterogeneous HPC infrastructures. This is achieved by combining
the operator parallelism\cite{chan2004algorithm} using MPI and batched
matrix-vector and matrix-matrix multiplications using GPUs.
In particular, we show that the algorithm enables the use of
Tensor Cores for accelerating GEMM, which is 
the most computationally expensive part in DMRG.
With this new development, we can reach an unprecedented accuracy (1 mH per metal) 
for the ground-state energy of a CAS(114e,73o) model\cite{li2019electronic,linkToFCIDUMPpclusters} of the P-cluster with a bond dimension $D=14000$.

The present algorithm can be readily combined with other techniques to
further improve the performance of \emph{ab initio} DMRG, such as SU(2) symmetry, 
real-space parallelization\cite{stoudenmire2013real}, and 
mixed-precision schemes\cite{tian2022mixed}. We are also interested in 
applying the present algorithm to fully relativistic DMRG\cite{
knecht2014communication,battaglia2018efficient,zhai2022comparison,li2023time}, 
where the symmetry is lower and hence
the symmetry blocks are much larger and complex algebra is
required. We expect in such case the GPU acceleration
will be even more beneficial.

\begin{acknowledgement}
We acknowledge Dingshun Lv, Qiaorui Chen, and Zhen Guo for helpful discussions on batched matrix-matrix multiplications on GPU, and Bingbing Suo for useful comments on the manuscript.
This work was supported by the National Natural Science Foundation of China (Grants No. 21973003)
and the Fundamental Research Funds for the Central Universities.
We acknowledge the computing resource and technical support provided by Tan Kah Kee Supercomputing Center (IKKEM).
\end{acknowledgement}


\bibliography{refSPMPS}

\end{document}